\documentclass[sigconf]{acmart}

\usepackage{pifont}
\usepackage{float}
\usepackage{multirow}

\AtBeginDocument{%
  }

\setcopyright{acmcopyright}
\copyrightyear{2018}
\acmYear{2018}
\acmDOI{XXXXXXX.XXXXXXX}

\acmConference[arXiv preprint]{arXiv.org e-Print archive}{arXiv.org}{e-Print archive}

\acmPrice{15.00}
\acmISBN{978-1-4503-XXXX-X/18/06}

\acmSubmissionID{596}

\renewcommand\footnotetextcopyrightpermission[1]{}
\begin{document}

\title{Fine-Grained Music Plagiarism Detection: \\
Revealing Plagiarists through Bipartite Graph Matching and a Comprehensive Large-Scale Dataset}

\author{Wenxuan Liu, Tianyao He, Chen Gong, Ning Zhang$^*$, Hua Yang$^*$, Junchi Yan}
\affiliation{%
  \institution{MOE Key Lab of Artificial Intelligence, Shanghai Jiao Tong University}
  \city{Shanghai}
  \country{China}
}
\email{{wenxuanliu,hetianyao,gongchen,ningz,hyang,yanjunchi}@sjtu.edu.cn}
\thanks{* Corresponding authors}

\renewcommand{\shortauthors}{Liu et al.}

\begin{abstract}
Music plagiarism detection is gaining more and more attention due to the popularity of music production and society's emphasis on intellectual property. We aim to find fine-grained plagiarism in music pairs since conventional methods are coarse-grained and cannot match real-life scenarios. Considering that there is no sizeable dataset designed for the music plagiarism task, we establish a large-scale simulated dataset, named Music Plagiarism Detection Dataset (MPD-Set) under the guidance and expertise of renowned researchers from national-level professional institutions in the field of music. MPD-Set considers diverse music plagiarism cases found in real life from the melodic, rhythmic, and tonal levels respectively. Further, we establish a Real-life Dataset for evaluation, where all plagiarism pairs are real cases. To detect the fine-grained plagiarism pairs effectively, we propose a graph-based method called Bipatite Melody Matching Detector (BMM-Det), which formulates the problem as a max matching problem in the bipartite graph. Experimental results on both the simulated and Real-life Datasets demonstrate that BMM-Det outperforms the existing plagiarism detection methods, and is robust to common plagiarism cases like transpositions, pitch shifts, duration variance, and melody change. Datasets and source code are open-sourced at \url{https://github.com/xuan301/BMMDet_MPDSet}.
\end{abstract}

\ccsdesc[500]{Applied computing~Sound and music computing}
\ccsdesc[300]{Theory of computation~Shortest paths}
\ccsdesc[500]{Information systems~Near-duplicate and plagiarism detection}

\keywords{music plagiarism, bipartite graph matching, edit distance, sequence similarity}


\maketitle

\section{Introduction and Related Work}
    Nowadays, the number of music documents on the Internet is increasing rapidly. Each year, over 10 million new albums of recorded music are released and over 100 million new musical pieces are registered for copyright~\cite{inproceedings}. Easy access to musical content increases the risk of unintentional or intentional plagiarism. The number of lawsuits and revenue losses due to music plagiarism is soaring~\cite{de2016visualization}. However, no concrete rules have seen proposed for music copyright infringement~\cite{de2017music}. 
    
    Melody plagiarism is prominent in the accusation, though sample and rhythm plagiarism are also common~\cite{dittmar2012audio}. The melody plagiarism problem has been researched a lot over the past decades \cite{de2017fuzzy,de2017computational, pachet2006cuidado}. Several methods have been proposed to extract melody from musical mixtures \cite{9053024, 7952144, 7178348, 9414190}.
    Existing plagiarism detection methods can be categorized as audio-based methods and sheet-based methods. 
    The audio-based methods inspect the music pairs by comparing the time-frequency representation of their audio~\cite{dittmar2012audio, downie2008audio, borkar2021music}. The sheet-based method measures the symbolic melodic similarity~\cite{siedetecting, velardo2016symbolic}. 
    However, the current methods are coarse-grained and consider little about music theory. They lack robustness and perform poorly when facing some tricky music changes like transposition, pitch shifts, duration variance, and melody change. Some high-level feature based methods like~\cite{robine2007adaptation,10.1145/1290082.1290103} formulate the problem as a sequence similarity problem solved by edit distance. Other works define the plagiarism degree based on n-gram techniques such as Ukkonen measure, Sum Common measure, and TF-IDF correlation \cite{doraisamy2003robust,bainbridge2005searching, doi:10.1177/102986490901300111}. Nevertheless, these methods cannot perform \textbf{fine-grained} detection, which means that  plagiarism only exists in a relatively small part of a musical piece, in the form of pitch shifts, duration variance, and melody change, etc.
    
    \begin{figure*}[tb!]
        \includegraphics[width=0.95\textwidth]{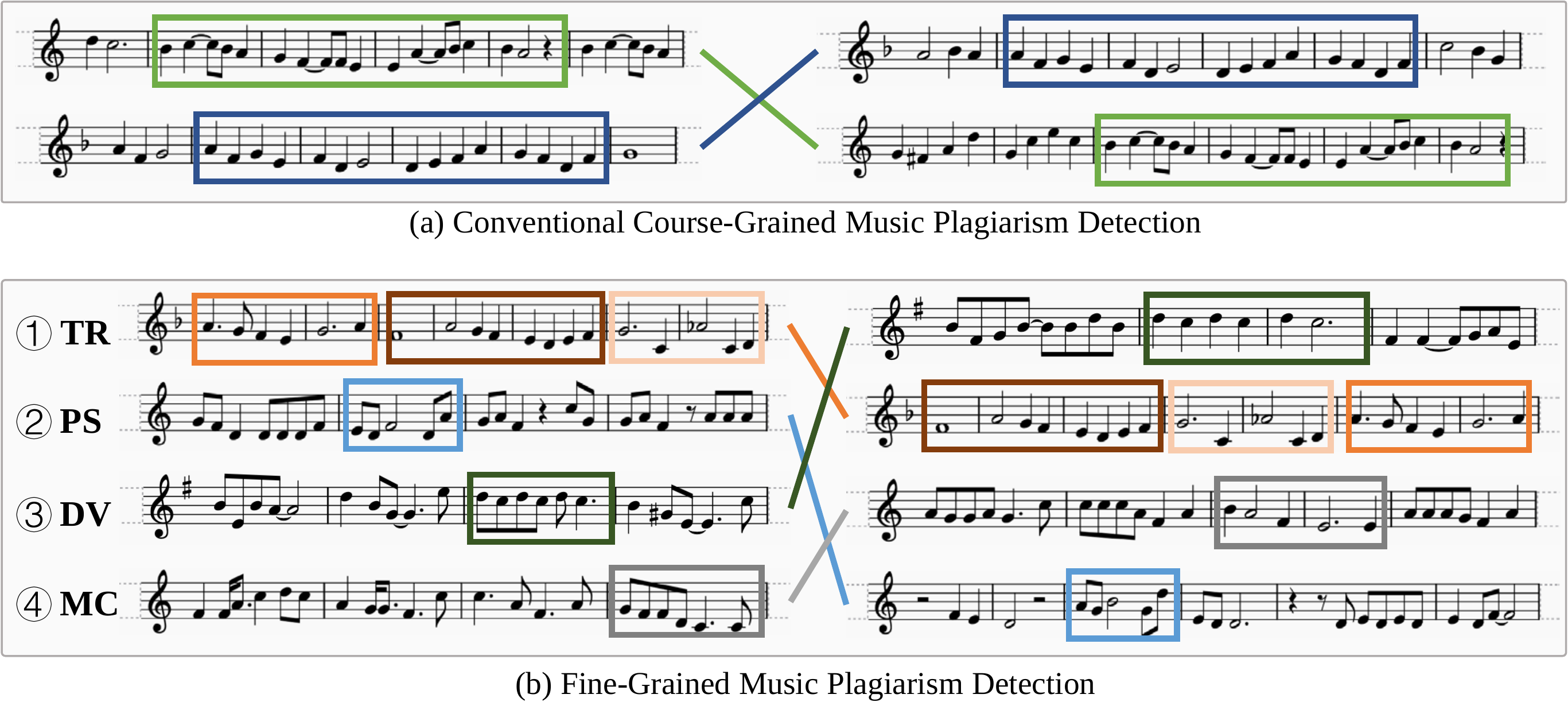}
        \caption{Comparison between the conventional and fine-grained music plagiarism detection is particularly studied in this paper. \textbf{(a)} Conventional plagiarism detection is usually coarse-grained and not optimized to identify common real-life plagiarism scenarios. \textbf{(b)} BMM-Det can perform fine-grained music plagiarism detection, accurately identify even when the proportion of plagiarised clips is low, and optimize the identification of plagiarism cases such as \ding{192} transposition (TR), \ding{193} pitch shifts (PS), \ding{194} duration variance (DV), \ding{195} melody change (MC), etc.}
        \label{shifting}
    \end{figure*}

    We propose a novel model, Bipartite Melody Matching Detector (BMM-Det), for fine-grained music plagiarism detection, which can find local melody plagiarism pairs from music datasets. Compared with conventional methods, BMM-Det is able to detect finer-grained plagiarised fragments and identify hard cases like transposition, duration variance, pitch shifts, melody change, etc. The detailed difference is shown in Figure \ref{shifting}. BMM-Det converts two melodies into a bipartite graph and regards the corresponding maximum weight matching as their plagiarism degree. Based on music theory, we represent each melody as a sequence and regard segments of the sequence as vertices. The edge's weight between two vertices is defined by an elaborately designed distance. In our dataset constructed from real-life cases, our method surpasses the baseline plagiarism detection algorithms by a large margin.
    
    Apart from the ineffectiveness of existing plagiarism detection algorithms, there are no large-scale datasets collected and established for music plagiarism detection. 
    Existing datasets on music plagiarism detection like \cite{doi:10.1177/102986490901300111}, Columbia Law School \cite{de2017fuzzy} and MIREX $\footnote{\url{https://www.music-ir.org/mirex/wiki/2005:Symbolic_Melodic}}$ are not public and also out-of-date. Thus, we collect real-life music plagiarism cases on our own, which have been made public for all research use. Also, there is no large-scale dataset on music plagiarism, and court decisions are scarce. Datasets like POP909~\cite{wang2020pop909} contain multiple versions of the same piece of music but are designed for music generation. These multiple versions of music do not take into account any similarities in musical structure, pitch, or duration, but rather are a stylistically consistent re-creation of the original piece. Musical plagiarism is usually a fine-grained copy of another piece, but this type of dataset contains an overall variation. Therefore, we propose Music Plagiarism Detection Dataset (MPD-Set) specifically designed for music plagiarism detection, which is the first large-scale music plagiarism detection dataset and will be covered in detail in Section~\ref{sec:plagset}.

    \textbf{Our contributions are summarized as follows:}
    
   1. A fine-grained music plagiarism detection model based on bipartite graph matching named as BMM-Det is proposed. BMM-Det can cope with  transposition, pitch shifts, duration variances, and melody change, etc. It can also pick out the fine-grained similar regions between two musical pieces with low global similarity.
   
   2. Two new datasets for  music plagiarism detection are published. MPD-Set is a large-scale dataset designed under the guidance of researchers from renowned national-level professional institutions in the field
of music, and we also collect a dataset consisting of real-life plagiarism pairs. These datasets address the current lack of data in this field and will facilitate research on music plagiarism detection.
   
   3. To evaluate the performance of BMM-Det, the model parameters have been tuned using the training part of MPD-Set, and then tested directly on the testing part of MPD-Set and the whole Real-life Dataset. The experimental results indicate that BMM-Det is efficient in detecting fine-grained plagiarism and that the MPD-Set is an effective reflection of real-life plagiarism scenarios, highlighting the importance of our research in contributing to fairness within the music industry's copyright landscape.

    \begin{figure*}
    \centering
        \includegraphics[width=0.9\textwidth]{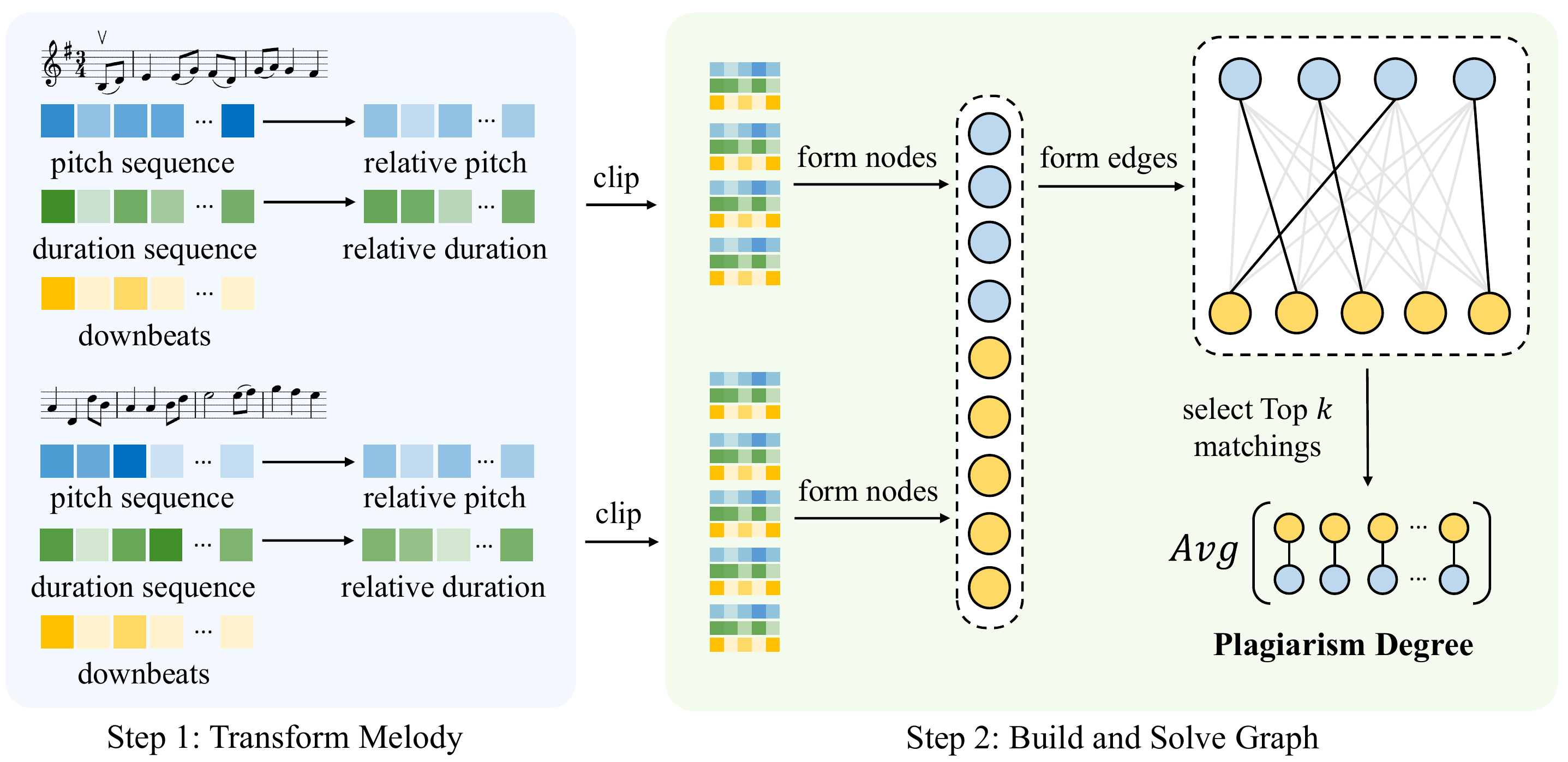}
        \caption{The pipeline of BMM-Det. Given two melodies, we convert them into two sequences, containing robust pitch, duration, and downbeat representations. Next, we cut the sequences into segments and build the bipartite graph. The maximum weight matching algorithm can be performed to get the final plagiarism degree and fine-grained matching results.}
        \label{fig:procedure}
    \end{figure*}
\section{The Simulated MPD-Set}\label{sec:plagset}
The Music Plagiarism Detection Dataset (MPD-Set) is, to the best of our knowledge, the first publicly available large-scale dataset encompassing 2,000 music pieces specifically designed for the task of music plagiarism detection. We have collaborated with researchers from renowned national-level professional institutions in the field of music to design the dataset under their guidance and expertise\footnote{Detailed 
information will be included in the final version of acknowledgements}. This joint effort ensures the MPD-Set accurately reflects the diverse range of music plagiarism cases encountered in real life, fostering the development of more effective tools and methods for protecting intellectual property in the music industry.


The original songs utilized to create the MPD-Set have been sourced from Wikifonia\footnote{\url{https://www.wikifonia.org}}, an open-source dataset comprised of real-life human-composed songs. To facilitate the construction of MPD-Set, we extract song fragments from Wikifonia in MusicXML format, convert them into the widely-used MIDI format within academia, and subsequently build the dataset.

The MPD-Set consists of a total of 2,000 music pieces, with each pair exhibiting a copying relationship. The most common method of plagiarism in the real world is often to select the most easily recognized elements (melody, rhythm, tonality) in music for direct replication or subtle revision. Therefore, to reflect real-world plagiarism occurrences and cover the most common plagiarism methods, we have designed four distinct types of plagiarism methods for the dataset from the melodic, rhythmic, and tonal levels respectively: transposition based on the melodic level, pitch shifts based on the tonal level, duration variance based on the rhythmic level, and melody change based on the melodic and rhythmic level, with each type accounting for 25\% of the dataset. The specific implications of these four plagiarism types are as follows:

\textbf{Transposition}: In the transposition type of plagiarism, the original song is randomly divided into 3 to 5 segments. After the order of these segments is shuffled, they are reassembled to create a new arrangement of the piece. This process results in a modified version of the original song with the overall structure altered.

\textbf{Pitch Shifts}: In this scenario, a fragment of the original song undergoes pitch shifts and is subsequently added to an entirely unrelated song, creating a new piece with embedded plagiarism. Although the position of the note sequence in the melodic fragment after shift is different from the previous one, the fragment of the original song are exactly the same melodic lines as the new piece, which is actually the presence of a melody in a different tonality. This type of plagiarism is to transfer the tonality of the original song melody, which is a relatively concealed and common method of musical plagiarism.

\textbf{Duration Variance}: This type of plagiarism varies from the process of pitch shifts, instead of shifting the pitch of the notes, the duration of each note in the original piece is altered, modifying the original song at the rhythmic level, and the manipulated fragment is added to a completely unrelated song. It is also a relatively concealed and common method of music plagiarism.

\textbf{Melody Change}: This type is a much more sophisticated situation of plagiarism, which entails simultaneously altering the melody and rhythm of a fragment of the original song using MuseMorphose~\cite{wu2021musemorphose}, a Transformer-based Variational Autoencoder (VAE) model, and integrating the transformed fragment into a completely unrelated song, resulting in a new piece containing concealed plagiarism. Concerning the melody change type of plagiarism, it is important to emphasize that, according to various national laws, determining whether a direct plagiarism relationship exists between the generated music and the original piece can be quite challenging. This difficulty arises because the altered melody and rhythm, despite being derived from the original song, may exhibit significant differences, making it hard to establish a clear connection between the two.

\section{Method}
\subsection{Preliminaries: Bipartite Graph Matching}
    Bipartite Graph is a type of graph where the nodes can be divided into two separate sets, often referred to as "left" and "right." Edges only exist between nodes that are in different sets (i.e. a node in the "left" set can only be connected to a node in the "right" set, and vice versa). Bipartite graph matching refers to the problem of finding pairs of nodes in a bipartite graph such that each node is paired with exactly one other node, and no two pairs share an edge. In other words, we want to find a way to connect every node on the left side of the graph with a node on the right side in such a way that there are no overlaps. 
    
    In mathematical formulation, Given a bipartite graph $G=(V,E)$ consisting of two disjoint sets of vertices $U$ and $W$, where $|U|=n$ and $|V|=m(n\geq m)$, and an edge set $E\subseteq U\times W$. The goal is to find the maximum cardinality matching $M$ in $G$, where $M\subseteq E$. Formally, let $M$ be a subset of edges that form a matching in $G$. We can represent $M$ as a binary vector $(x_1, x_2, \dots, x_m)$, where $x_i=1$ if edge $i$ belongs to $M$, and $x_i=0$ otherwise. The cost of each edge is $S=(s_1, s_2, \dots, s_m)$. Then, we can formulate it as an optimization problem:
    \begin{equation}
    \renewcommand{\arraystretch}{1.2}
    \begin{array}{rll}
        \mathrm{maximize} & \sum_{i=1}^{m} s_i x_i & \\
        s.t. & x_i\in \{0,1\} & \forall i\in\{1,2,\dots, m\} \\ 
             & \sum_{i\in N(j)}x_i\leq 1 & \forall i\in\{1,2,\dots, n\} \\
    \end{array}
    \end{equation}
    where $N(j)$ denotes the set of nodes adjacent to node $j$ in the bipartite graph $G$. The first constraint ensures that every edge is either selected or not selected in the matching, and the second constraint ensures that each vertex in $U$ is adjacent to at most one vertex in $W$ in the matching.
    
\subsection{Melody Sequence Representation}\label{sec:feature}
    Before the detection algorithm, we first process the input data based on music theory. Specifically, given two musical pieces, we represent their melodies with two sequences $S_1=\{a_1,\dots,a_{n}\}$ and $S_2=\{b_1,\dots,b_m\}$. Every component $a_i$ is composed of the \textit{pitch}, \textit{duration}, and \textit{downbeat} of the corresponding note. The \textit{pitch} and \textit{duration} are denoted in Musical Instrument Digital Interface(MIDI) protocol, and the binary value \textit{downbeat} denotes whether this note is downbeat.
    
    The changes of key and speed in music are commonly inspected in music plagiarism. To handle the challenges, we apply the idea of the relative sequence for both \textit{pitch} and \textit{duration}. The relative sequences record the pitch and duration difference between neighboring notes instead of the absolute pitch, which makes our melody representation robust to key and speed modifications.
    
\subsection{Music Plagiarism Detection with Bipartite Graph Matching}\label{sec:bipartite}
   With the melody representation, we can formulate music plagiarism detection as a bipartite graph matching problem. We establish a bipartite graph $G=(U\cup W, E)$, where $U$ and $W$ are two disjoint vertex sets of the bipartite graph with $|U|=n, |W|=m$, and $E$ represents the edge set between $U$ and $W$. 

\subsubsection{\textbf{Vertex formation}}
    We choose to get the vertices from the melody sequence by cutting with overlaps. For two melody sequences $S_1$ and $S_2$, we cut them into clips with length $l$ and overlapping rate $r$ to obtain clip lists $P_1$ and $P_2$. Each clip in $P_1$ is considered as a vertex in $U$ and each clip in $P_2$ is a vertex in $W$.

\subsubsection{\textbf{Edge formation}}
    For each vertex $u\in U$ and $w\in W$, we construct an edge $e=(u,w)$. In music plagiarism, high melody similarity is a determinant in the final judgment, while low similarity is acceptable. Therefore, we want our edge cost to satisfy two properties. \textbf{First}, the edge cost should reflect the similarity of two melody clips. \textbf{Second}, the edge cost should amplify the high similarity and suppress the low similarity, making it more sensitive to plagiarism. 
    
    To satisfy the first property, we use the edit distance to reflect the melody similarity. The edit distance between two sequences aims to calculate the minimum operation costs to change one sequence to the other with pre-defined operations. The three basic operations on melody are substitution, insertion, and deletion, whose costs $c_{\mathrm{sub}}$, $c_{\mathrm{ins}}$ and $c_{\mathrm{del}}$ are detailed in section \ref{sec:cost_define}.
    
    The edit distance can be calculated with the idea of dynamic programming. We define a two-dimensional dynamic table $d$, where $d_{i,j}$ records the distance between the first $i$ notes of $u$ and the first $j$ notes of $v$. Therefore, $d_{l,l}$ means the complete edit distance between $u$ and $w$. We can define the boundary conditions:
    \begin{equation}
        d_{i,0} = \sum_{k=1}^{i} c_{\mathrm{del}}(u_k), 1\leq i\leq l
    \end{equation}
    
    \begin{equation}
        d_{0,j} = \sum_{k=1}^{j} c_{\mathrm{ins}}(w_k), 1\leq j\leq l
    \end{equation}
    where $u_k$ means the $k_{\mathrm{th}}$ note of vertex $u$ (similar for $w$).
    With the boundary conditions, we can define the update function under different circumstances. 
    
    For $u_i=w_j$, we have:
    \begin{equation}
        d_{i,j} = d_{i-1,j-1}
    \end{equation}
    
    For $u_i\neq w_j$, we have:
    \begin{equation}
        d_{i,j} =\min(d_{i,j}^{\mathrm{del}},\ d_{i,j}^{\mathrm{ins}},\ d_{i,j}^{\mathrm{sub}})
    \end{equation}
    where
    \begin{equation}
        \begin{aligned}
            d_{i,j}^{\mathrm{del}} &= d_{i-1,j}+c_{\mathrm{del}}(u_i) \\
            d_{i,j}^{\mathrm{ins}} &= d_{i,j-1}+c_{\mathrm{ins}}(w_j) \\
            d_{i,j}^{\mathrm{sub}} &= d_{i-1,j-1}+c_{\mathrm{sub}}(u_i,w_j) \\
        \end{aligned}
    \end{equation}
    
    
    Now that we have obtained the edge weight between two nodes, we aim to redesign it to better suit the needs of music plagiarism detection. We observe that in music plagiarism, the local high similarity of two short clips often contributes more to the judgment, while low similarity is acceptable in music creation. To achieve this, we want to satisfy two properties for our edge weight:

\textbf{1)} Amplify high similarity and suppress low similarity.

\textbf{2)} Ensure the edge weight falls within the range of $[0, 1]$.

To satisfy the second property, we use the function described in Equation \ref{equ:trans_func} to transform the original edit distance into our required edge weight. The logarithm calculation in Equation \ref{equ:trans_func} is used to amplify the differences between smaller values and suppress those between larger values.
        \begin{equation}
        f(d)=\frac{\ln(1+e^{-d})}{\ln 2}
        \label{equ:trans_func}
    \end{equation}
\subsubsection{\textbf{Music Plagiarism Degree}}

With the bipartite graph established, we can now analyze the comprehensive melody similarity between two music pieces with fine-grained awareness. To obtain the minimum cost matching of the bipartite graph $G$, we employ the Kuhn–Munkres (KM) algorithm~\cite{https://doi.org/10.1002/nav.3800020109}. The KM algorithm, also known as the Hungarian algorithm, is a combinatorial optimization algorithm that efficiently solves the assignment problem by computing the minimum weight of matching in a weighted bipartite graph.

Let $G = (U \cup W, E)$ be a weighted bipartite graph, where $U$ and $W$ are the sets of vertices on the two sides of the graph, and $E$ is the set of edges connecting the vertices. Each edge $(u, w) \in E$ has an associated non-negative weight $t(u, w)$. The goal of the KM algorithm is to find a perfect matching $M$ of the bipartite graph that minimizes the total weight:

\begin{equation}
M = \arg\min_{M} \sum_{(u, w) \in M} t(u, w).
\end{equation}

The KM algorithm finds the minimum weight perfect matching by iteratively updating a set of labels, one for each vertex in $U$ and $W$. The algorithm starts with an initial feasible labeling and a partial matching. It then searches for augmenting paths, which are alternating paths that start and end with unmatched vertices, while respecting the labeling constraints. If an augmenting path is found, the algorithm increases the size of the matching by flipping the matched and unmatched edges along the path. The process repeats until no more augmenting paths can be found, at which point the algorithm has found a minimum weight perfect matching.

By applying the KM algorithm, we can determine the minimum matching scores, which represent the plagiarism degree between the two music pieces. Furthermore, we can locate the plagiarized parts by examining the matched vertex pairs in the bipartite graph. This approach allows for a detailed analysis of the similarities between the two music pieces, identifying and quantifying the extent of plagiarism present in the compositions.
    
\subsection{Score Design of Edit Operations}
\label{sec:cost_define}
    In this section, we define the scores of substitution, insertion, and deletion. For insertion and deletion, we set their costs to $c_{\mathrm{ins}}=c_{\mathrm{del}}=1$. For substitution, the computation is more complicated. The common knowledge is that substitution with larger pitch or duration differences will have a greater influence on the music, especially the downbeat notes. Therefore, we need a comprehensive design of the substitution cost to reflect its real impact on the music.
    
    Our substitution cost function is Equation \ref{equ:total_cost}:
    \begin{equation}
        c_{\mathrm{sub}}(u_i, w_j)=c_{\mathrm{downbeat}}(u_i, w_j)\cdot [c_{\mathrm{pitch}}(u_i, w_j) + c_{\mathrm{duration}}(u_i, w_j)],
        \label{equ:total_cost}
    \end{equation}
    where $c_{\mathrm{downbeat}}$ is the downbeat coefficient, $c_{\mathrm{pitch}}$ is the pitch difference cost, and $c_{\mathrm{duration}}$ is the duration difference cost. $c_{\mathrm{pitch}}$ and $c_{\mathrm{duration}}$ are defined as:
    \begin{equation}
        c_{\mathrm{pitch}}(u_i, w_j)=\left|u_{i}^{\mathrm{pitch}}-w_{j}^{\mathrm{pitch}}\right|
        \label{equ:pitch_cost}
    \end{equation}
    
    \begin{equation}
        c_{\mathrm{duration}}(u_i, w_j)=\left|u_{i}^{\mathrm{duration}}-w_{j}^{\mathrm{duration}}\right|
        \label{equ:duration_cost}
    \end{equation}
    where the superscript represents that it is the pitch or duration of the note. Downbeat coefficient $c_{\mathrm{downbeat}}$ is defined in Equation \ref{equ:downbeat_coe}:
    \begin{equation}
        c_{\mathrm{downbeat}}(u_i, w_j)=k_{\mathrm{down}}\cdot u_i^{\mathrm{downbeat}}\cdot w_j^{\mathrm{downbeat}}
        \label{equ:downbeat_coe}
    \end{equation}
    where $u_i^{\mathrm{downbeat}}\in\{0,1\}$ is the binary downbeat indicator representing whether the $i_{th}$ note of $u$ is a downbeat(similar for $v$). $k_{\mathrm{down}}$ is a hyperparameter controlling the importance of the downbeat.
    
    
    
    
\section{Experiments}\label{sec:experiment}
    To demonstrate the effectiveness of our MPD-Set and BMM-Det for fine-grained plagiarism detection, we opt to tune BMM-Det using the MPD-Set and evaluate its performance on two test sets. The first test set is a subset of the MPD-Set, and the second test set consists of real-life plagiarism cases (Real-life Dataset). This approach allows us to assess the algorithm's ability to generalize across different types of data, showing the robustness of our method.
    
\subsection{Datasets}
    We use two datasets for the experiment. The first is MDP-Set, which has been introduced in Section~\ref{sec:plagset}. The second real-life one is collected by us and consists of 29 pairs of songs, where 20 pairs are legally judged as plagiarism by the court~\cite{musicCourt}, and the other 9 pairs of songs are from Ping An Tech's work$\footnote{\url{https://github.com/andyjhj/MusicPlag\_Demo}}$, and all these 29 pairs of songs constitute plagiarism. 

    

\subsection{Implementation Details}
    \begin{figure*}[tb!]
        \centering
        \includegraphics[width=0.9\textwidth]{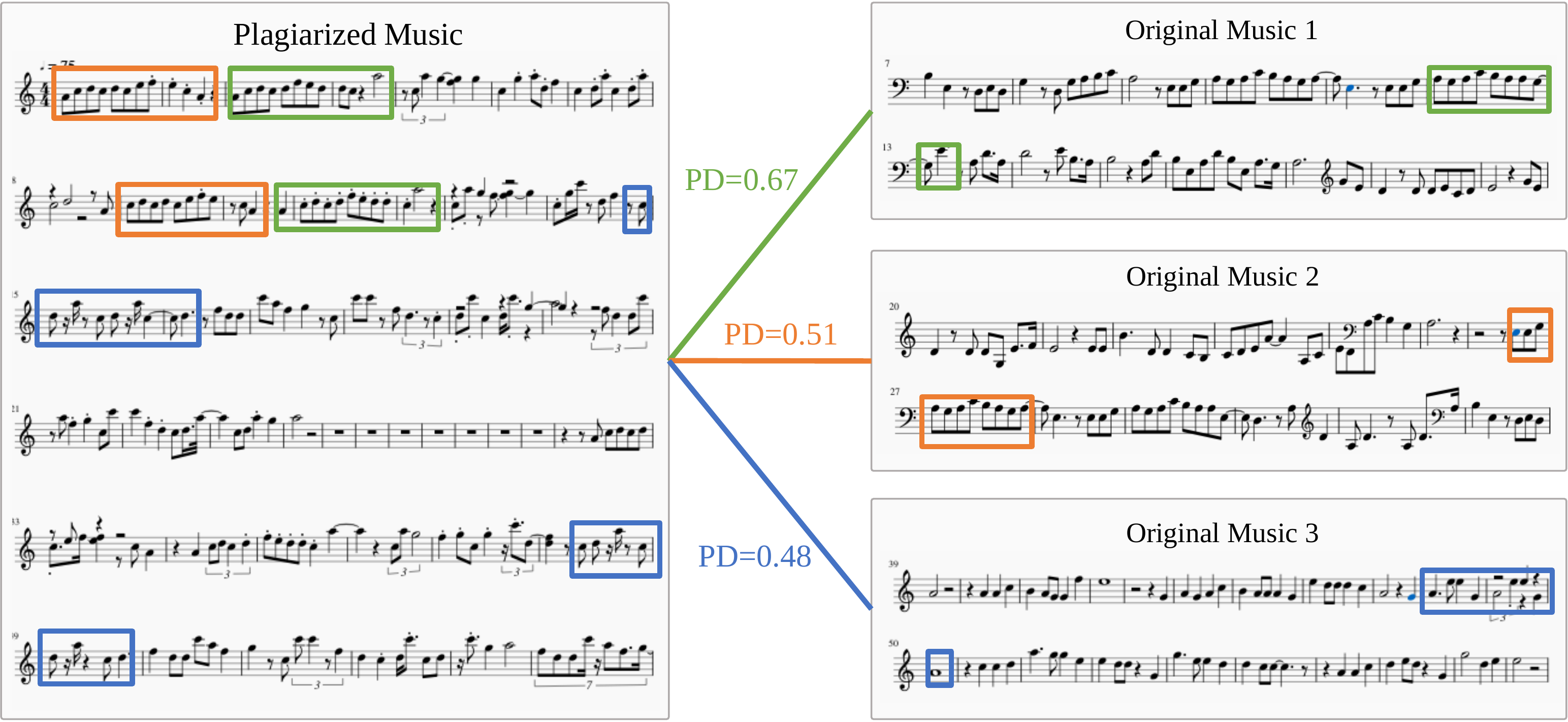}
        \caption{An example of fine-grained plagiarism detection using BMM-Det. The left song has potential plagiarism compared to the three original songs on the right. Musical pieces of the same colour are detected as plagiarised pairs with high plagiarism scores (e.g., 0.67, 0.51, 0.48), indicating a high degree of similarity between the compared sections.}
        \label{fig:demo}
    \end{figure*}
\textbf{Experiment Setting: }
Our primary objective is to assess the capability of our algorithm to detect music plagiarism and locate plagiarized parts, especially in real-world situations. Due to the limited availability of real-life music plagiarism legal cases, we strive to significantly expand our test sets to showcase the robustness and adaptability of our model. To achieve this, we first divide the MPD-Set into an 80-20 proportion. We use 80\% of the MPD-Set, consisting of approximately 1600 songs, to tune the BMM-Det algorithm. The remaining 20\% of the MPD-Set, comprising approximately 400 songs, serves as the first test set. In addition, we utilize a second test set consisting of real-life plagiarism cases to further evaluate our model's performance in practical scenarios.

By tuning the algorithm on the majority of the MPD-Set and striving to achieve impressive results on the Real-life Dataset, we aim not only to demonstrate the effectiveness of our BMM-Det model in capturing and reflecting the characteristics of plagiarism present in real-world scenarios, but also to highlight the importance of the MPD-Set as a reliable representation of real-life plagiarism cases. Through this approach, we intend to evaluate the algorithm's ability to adapt and generalize across various types of data, while ensuring a comprehensive assessment of its performance in detecting  music plagiarism.

\textbf{Evaluation Metrics: }
To measure the performance of our algorithm, we mainly focus on two indicators: the Average Ranking Index (ARI) of the plagiarized songs and the accuracy.

1. \textbf{Average Ranking Index (ARI)}: The ARI measures the average ranking of the plagiarized song within a dataset when compared to the remaining original songs. The lower the ARI, the better the performance of the algorithm. Mathematically, it is defined as:

\begin{equation}
    ARI = \frac{1}{N} \sum_{i=1}^{N} R_{i},
\end{equation}

where $N$ is the number of song pairs in the test set and $R_{i}$ is the ranking of the plagiarized song for the $i$-th test case.

2. \textbf{Accuracy}: Accuracy measures the proportion of correctly identified plagiarized songs in the test set. The higher the accuracy, the better the performance of the algorithm. It is defined as:

\begin{equation}
    \text{Accuracy} = \frac{\text{Number of Correct Identifications}}{\text{Total Number of Test Cases}} \times 100\%.
\end{equation}

In our experiment, a result is considered correct if the plagiarized song is ranked first when compared to the other songs in the dataset. By evaluating these two metrics, we can obtain a comprehensive understanding of our algorithm's performance in detecting music plagiarism.



\subsection{Results and Analysis}
    \textbf{On Testing Dataset.} We compare BMM-Det with other existing methods, including Sum Common~\cite{0cognitive}, Ukkonen~\cite{0cognitive}, TF-IDF correlation~\cite{2010speech}, and Tversky-equal~\cite{1988Features}, on both the MPD-Set and the Real-life Dataset. Ukkonen and Sum Common consider the difference of all n-gram features occurring in either one musical piece~\cite{0cognitive}.
    TF-IDF correlation method is widely used for retrieving text documents~\cite{2010speech}. In our experiment, we use n-gram features weighted by their frequency in both musical pieces and our dataset, which is measured by inverted document frequency~\cite{1999Foundations}: $\mathrm{IDF}(\tau)=\log(\frac{n}{n_{\tau}})$ where $n$ is the size of the dataset and $n_{\tau}$ is the number of pieces including $\tau$. Tversky's ratio model originates from~\cite{1988Features}, which is adapted by inserting $\mathrm{IDF}$.
    \begin{table}[tb!]
  \centering
  \caption{Comparison of average index and accuracy between different methods on the MPD-Set and the Real-life Dataset.}
    \begin{tabular}{c|c|c|c|c}
    \toprule
          & \multicolumn{2}{c|}{MPD-Set} & \multicolumn{2}{c}{Real-life Dataset} \\
          & Avg. Index  & Acc.  & Avg. Index  & Acc. \\
    \midrule
    TF-IDF corr.\cite{2010speech} & 10.52 & 0.695 & 3.24  & 0.655 \\
    Tversky-equal\cite{1988Features} & 16.37 & 0.685 & 3.14  & 0.655 \\
    Sum Common\cite{0cognitive} & 12.00    & 0.653 & 3.10   & 0.724 \\
    Ukkonen\cite{0cognitive} & 10.08 & 0.675 & 3.00     & 0.759 \\
    \midrule
    BMM-Det (Ours) & \textbf{7.07}  & \textbf{0.705} & \textbf{2.17}   & \textbf{0.828} \\
    \bottomrule
    \end{tabular}%
  \label{tab:compare_method_2}%
\end{table}%

    The results are shown in Table~\ref{tab:compare_method_2}. For each baseline, we tune the hyperparameter $n$. It is evident that BMM-Det outperforms all the baselines on both datasets. This indicates that methods based on n-gram features work only when the frequency of common n-gram features is related to the overall plagiarism degree. Upon further analysis of the results in the table, we can observe the following:
    
    1. On the MPD-Set, BMM-Det (our method) outperforms other methods with an average index of 7.07 and an accuracy of 0.705. This is a significant improvement compared to the second-best method, Ukkonen, which has an average index of 10.08 and an accuracy of 0.675. This demonstrates the effectiveness of BMM-Det in detecting  music plagiarism in the MPD-Set.
    
    2. On the Real-life Dataset, BMM-Det also achieves the best performance with an average index of 2.17 and an accuracy of 0.828. The second-best method, Ukkonen, has an average index of 3.00 and an accuracy of 0.759. The superior performance of BMM-Det on the Real-life Dataset further confirms its capability to adapt and generalize across various types of data, including real-life plagiarism cases.
    
    3. Comparing the results between MPD-Set and the Real-life Dataset, BMM-Det maintains consistently high performance in both datasets. This highlights the importance of the MPD-Set as a reliable representation of real-life plagiarism cases and validates the effectiveness of our algorithm in capturing and reflecting the characteristics of plagiarism present in real-world scenarios. The performance of BMM-Det can be attributed to its ability to adapt and generalize across different types of data, making it a more robust and reliable method for detecting  music plagiarism.
    
    In conclusion, the results of our experiment demonstrate the effectiveness of the BMM-Det algorithm in detecting  music plagiarism in various testing datasets, including real-life cases. Its superior performance over other methods, both in terms of the average index and accuracy, highlights its potential for practical applications in the field of music plagiarism detection.

     \begin{table*}[t!]
         \centering
         \caption{Ablation studies on different compositions of optimization methods for plagiarism degree measurement on MPD-Set and the Real-life Dataset. Here RelativeDuration means using relative duration sequence, RelativePitch means using relative pitch sequence, MaxMatch means segmenting the sequence and computing maximum weight matching, DownBeat means considering whether the note is downbeat, and NoteDistance means considering pitch shifts when computing edit distance.}
\resizebox{.98\linewidth}{!}{
        \begin{tabular}{ccccc|c|c|c|c}
        \toprule
        \multicolumn{1}{c}{\multirow{2}[0]{*}{RelativeDuration}} &
            \multicolumn{1}{c}{\multirow{2}[0]{*}{\centering RelativePitch}} &
            \multicolumn{1}{c}{\multirow{2}[0]{*}{\centering MaxMatch}} &
            \multicolumn{1}{c}{\multirow{2}[0]{*}{\centering DownBeat}} &
            \multicolumn{1}{c}{\multirow{2}[0]{*}{\centering NoteDistance}} &
            \multicolumn{2}{|c|}{MPD-Set} & \multicolumn{2}{c}{Real-life Dataset} \\
          &       &       &       &       &  Avg. Index   & Acc.   &  Avg. Index   & Acc. \\
        \midrule
         \checkmark & \checkmark & & & & 36.32&0.365& 3.24& 0.482\\
         \checkmark &  & \checkmark &   &   & 32.26    & 0.435 & 5.00& 0.586\\
         \checkmark & & \checkmark & \checkmark &   & 32.04 & 0.445 & 5.17& 0.586\\
          & \checkmark & \checkmark &   & \checkmark & 10.01   & 0.665& 2.31& 0.689\\ 
          & \checkmark & \checkmark & \checkmark & \checkmark & 10.65 & 0.650& 2.28& 0.690\\ 
         \checkmark & \checkmark & \checkmark &   &   & 9.35& 0.635& 2.93& 0.689\\ 
         \checkmark & \checkmark & \checkmark & \checkmark &   & 9.00 & 0.630& 2.79& 0.758\\
         \midrule
         \checkmark & \checkmark & \checkmark &   & \checkmark & 7.07   & \textbf{0.705}& 2.17& \textbf{0.828}\\ 
         \checkmark & \checkmark & \checkmark & \checkmark & \checkmark & \textbf{6.98} & 0.700& \textbf{2.14}& \textbf{0.828}\\ 
        \bottomrule
        \end{tabular}}
        \label{tab:compare}

    \end{table*}

    \begin{table*}[tb!]
  \centering
  \caption{Comparison of Music Plagiarism Detection Methods Performance Across Different Plagiarism Types Using the MPD-Set Dataset}

    \begin{tabular}{c|c|c|c|c|c|c|c|c}
      \toprule
          & \multicolumn{2}{c|}{Transpostion} & \multicolumn{2}{c|}{Pitch Shifts} & \multicolumn{2}{c|}{Duration Variance} & \multicolumn{2}{c}{Melody Change} \\
          & Avg. Index & Acc.  & Avg. Index & Acc.  & Avg. Index & Acc.  & Avg. Index & Acc. \\
         \midrule
    TF-IDF corr.~\cite{2010speech} & 1.00  & 1.00  & 1.16  & 0.92  & 1.12  & 0.94  & 14.54  & 0.08  \\
    Tversky-equal~\cite{1988Features} & 1.00  & 1.00  & 1.34  & 0.98  & 1.06  & 0.96  & 13.92  & 0.14  \\
    Sum Common~\cite{0cognitive} & 1.00  & 1.00  & 2.16  & 0.82  & 1.14  & 0.90   & 17.02  & 0.04  \\
    Ukkonen~\cite{0cognitive} & 1.00  & 1.00  & 2.06  & 0.86  & 1.20  & 0.90  & 12.44  & 0.06  \\
    BMM-Det (Ours) & \textbf{1.00 } & \textbf{1.00 } & \textbf{1.02 } & \textbf{0.98 } & \textbf{1.00 } & \textbf{1.00 } & \textbf{6.89 } & \textbf{0.33 } \\
    \bottomrule
    \end{tabular}%
  \label{tab:testset}%
\end{table*}%

\textbf{Ablation Study:} In our experiment, we conducted ablation studies to test the impact of different combinations of optimization methods for plagiarism degree measurement on the MPD-Set and the Real-life Dataset. The results are shown in Table~\ref{tab:compare}. The tested optimization methods include RelativeDuration, RelativePitch, MaxMatch, DownBeat, and NoteDistance, where RelativeDuration means using relative duration sequence, RelativePitch means using relative pitch sequence, MaxMatch means segmenting the sequence and computing the maximum weight matching, Downbeat means considering whether the note is downbeat, and NoteDistance means considering the shifts of the pitch when computing the edit distance. 

From the table, we can draw the following observations:

1. When we analyze the results without MaxMatch, using only RelativeDuration and RelativePitch, we observe a decline in performance compared to when MaxMatch is included. In this case, the Average Ranking Index (ARI) on MPD-Set increases to 36.32, and the accuracy drops to 0.365. On the Real-life Dataset, the ARI increases to 3.24, and the accuracy decreases to 0.482. This result demonstrates the importance of MaxMatch as a key optimization method in our algorithm. MaxMatch is responsible for segmenting the sequence and computing the maximum weight matching, which significantly contributes to the robustness of the algorithm. Without MaxMatch, the algorithm becomes less effective at detecting  music plagiarism. 

2. When combining RelativeDuration, RelativePitch, and MaxMatch, we see a significant improvement in performance. The Average Ranking Index (ARI) on MPD-Set decreases to 9.35, and the accuracy increases to 0.635. On the Real-life Dataset, the ARI decreases to 2.93, and the accuracy increases to 0.689. This suggests that considering these three features together is essential for achieving better performance.

3. Adding more optimization methods such as DownBeat and NoteDistance to the combination of RelativeDuration, RelativePitch, and MaxMatch further improves the performance. The ARI on MPD-Set decreases to 6.98, and the accuracy increases to 0.700. On the Real-life Dataset, the ARI decreases to 2.14, and the accuracy increases to 0.828. This confirms that each optimization method contributes to the overall performance, and using them all together allows the algorithm to perform better in detecting  music plagiarism.

In conclusion, the ablation study demonstrates the effectiveness of combining optimization methods such as RelativeDuration, RelativePitch, MaxMatch, DownBeat, and Notedistance in detecting  music plagiarism. By considering multiple musical features, our algorithm becomes more robust, leading to improved performance on both MPD-Set and Real-life Datasets. This indicates that our BMM-Det algorithm effectively captures and reflects the characteristics of plagiarism present in real-world scenarios and highlights the importance of the MPD-Set as a reliable representation of real-life plagiarism cases.




\textbf{Considering Different Types of Plagiarism. }
In this experiment, our primary objective is to assess the effectiveness of various music plagiarism detection methods, including our proposed BMM-Det algorithm, across different types of plagiarism. The MPD-Set takes into account a comprehensive range of real-world plagiarism scenarios, making it an ideal choice for evaluating the performance of these methods. The dataset encompasses four distinct types of plagiarism: transposition, pitch shifts, duration variance, and melody change, with each type constituting 25\% of the dataset. By evaluating the performance of different methods in detecting  music plagiarism across these four types, we aim to gain a deeper understanding of their effectiveness in various real-life plagiarism scenarios.  The experimental results, presented in Table~\ref{tab:testset}, showcase the performance of different methods.

1. In the Transposition category, all methods achieve an average index of 1.00 and an accuracy of 1.00. This indicates that plagiarism involving transposition is relatively easy to detect and locate, as all methods perform equally well in this category.

2. For both Pitch Shifts and Duration Variance categories, BMM-Det outperforms all other methods. In the Pitch Shifts category, BMM-Det achieves the highest accuracy of 1.00 and an average index of 1.00. In the Duration Variance category, BMM-Det maintains a consistently high performance, achieving an average index of 1.02 and an accuracy of 0.98. These results demonstrate the superior performance of BMM-Det in detecting  music plagiarism involving pitch shifts and duration changes, highlighting its potential for practical applications in these specific types of plagiarism detection.

3. In the Melody Change category, BMM-Det significantly outperforms other methods, achieving an average index of 6.89 and an accuracy of 0.33. Although the accuracy is lower compared to other categories of plagiarism, it still demonstrates the potential of BMM-Det in detecting music plagiarism involving melody alterations. The lower accuracy may be attributed to the difficulty in establishing a clear connection between the generated music and the original piece due to significant differences in the altered melody.

The results of our experiment underscore the effectiveness of the BMM-Det algorithm in detecting music plagiarism across various types, including transposition, pitch shifts, duration variance, and melody change. Its superior performance in pitch shifts and duration variance categories highlights its potential for practical applications in the field of music plagiarism detection. The results also emphasize the importance of designing a dataset, such as the MPD-Set, that accurately reflects real-world plagiarism occurrences and challenges faced in detecting such cases.

\textbf{Identifying Fine-grained Plagiarized Parts:} BMM-Det is capable of identifying fine-grained plagiarized sections in music. Figure~\ref{fig:demo} displays qualitative results where a song is compared with others for potential plagiarism. The same color in the figure denotes the fine-grained plagiarized parts detected by BMM-Det. Upon listening to these parts, we find they exhibit a high degree of auditory similarity. Despite significant differences in structure, duration, and pitch between these pairs, BMM-Det can successfully detect the plagiarized sections, highlighting our method's robust capabilities.

In our approach, we process and compare potential plagiarized and original songs, calculating the edit distance using the BMM-Det algorithm. By analyzing these distances, we obtain the plagiarism scores (e.g., 0.67, 0.51, 0.48), which are derived from the maximum weight matching of segmented sequences in the compared songs. These scores indicate the degree of similarity between the compared song sections, serving as a valuable indicator to identify and assess the fine-grained plagiarized parts in the analyzed music. This approach demonstrates the effectiveness of BMM-Det in detecting  music plagiarism at a detailed level, as it effectively captures the characteristics of plagiarism present in real-world scenarios.

It is crucial to note that the determination of music plagiarism ultimately falls within the realm of legal judgment. Our BMM-Det algorithm serves as an auxiliary tool for music plagiarism detection. Users can decide on the plagiarism threshold according to their needs. Once the threshold is determined, users can delve into the results provided by BMM-Det to examine the detected plagiarized sections at a fine-grained level. This flexibility ensures that BMM-Det can cater to different requirements and preferences while effectively identifying music plagiarism.

\section{Conclusion and Discussion}
    
    Music plagiarism is a widespread issue in the music industry and has become increasingly difficult to detect due to the use of digital tools and the prevalence of online music distribution platforms. To tackle this problem, we propose BMM-Det that can effectively detect fine-grained music plagiarism across different datasets. BMM-Det is based on a bipartite graph and is robust to various forms of musical manipulation that are often used to disguise plagiarized music, such as transposition, duration variance, pitch shifts, and melody change. To evaluate the effectiveness of BMM-Det, we created a simulated large-scale dataset called MPD-Set, incorporating a range of different types of plagiarism generated using a specialized method designed to replicate real-world examples of music plagiarism. Furthermore, we collect a Real-life Dataset encompassing numerous real-life cases of music plagiarism. Experimental results on both MPD-Set and the Real-life Dataset demonstrate BMM-Det's outstanding performance in detecting fine-grained plagiarism. By developing BMM-Det for identifying fine-grained plagiarism and creating MPD-Set that accurately represents real-world plagiarism scenarios, our work contributes to promoting fairness and transparency in the music industry.
    
    We have made efforts to design our method as a transparent "white-box" approach, but we acknowledge that the final determination of music plagiarism is ultimately a legal matter. Therefore, we envision our method as a valuable auxiliary tool for music industry professionals to use for self-evaluation or detecting plagiarism in others' works. BMM-Det can provide detailed information on the fine-grained plagiarized segments and the degree of plagiarism, and its detection process is theoretically well-grounded, making it highly useful for related investigations.

For future work, we plan to improve BMM-Det by considering additional auditory features to enhance its accuracy and adaptability. Furthermore, we aim to expand our dataset to facilitate ongoing research in this field. We also encourage the research community to join our efforts in continually refining and expanding the dataset. Collaborative contributions will help in creating a more robust and comprehensive dataset that accurately reflects the diverse range of music plagiarism cases encountered in real life.


\bibliographystyle{ACM-Reference-Format}
\bibliography{sample-base}


\begin{thebibliography}{28}


\ifx \showCODEN    \undefined \def \showCODEN     #1{\unskip}     \fi
\ifx \showDOI      \undefined \def \showDOI       #1{#1}\fi
\ifx \showISBNx    \undefined \def \showISBNx     #1{\unskip}     \fi
\ifx \showISBNxiii \undefined \def \showISBNxiii  #1{\unskip}     \fi
\ifx \showISSN     \undefined \def \showISSN      #1{\unskip}     \fi
\ifx \showLCCN     \undefined \def \showLCCN      #1{\unskip}     \fi
\ifx \shownote     \undefined \def \shownote      #1{#1}          \fi
\ifx \showarticletitle \undefined \def \showarticletitle #1{#1}   \fi
\ifx \showURL      \undefined \def \showURL       {\relax}        \fi
\providecommand\bibfield[2]{#2}
\providecommand\bibinfo[2]{#2}
\providecommand\natexlab[1]{#1}
\providecommand\showeprint[2][]{arXiv:#2}

\bibitem[Bainbridge et~al\mbox{.}(2005)]%
        {bainbridge2005searching}
\bibfield{author}{\bibinfo{person}{David Bainbridge}, \bibinfo{person}{Michael
  Dewsnip}, {and} \bibinfo{person}{Ian~H Witten}.}
  \bibinfo{year}{2005}\natexlab{}.
\newblock \showarticletitle{Searching digital music libraries}.
\newblock \bibinfo{journal}{\emph{Information processing \& management}}
  \bibinfo{volume}{41}, \bibinfo{number}{1} (\bibinfo{year}{2005}),
  \bibinfo{pages}{41--56}.
\newblock


\bibitem[Bittner et~al\mbox{.}(2017)]%
        {7952144}
\bibfield{author}{\bibinfo{person}{Rachel~M. Bittner}, \bibinfo{person}{Avery
  Wang}, {and} \bibinfo{person}{Juan~P. Bello}.}
  \bibinfo{year}{2017}\natexlab{}.
\newblock \showarticletitle{Pitch contour tracking in music using Harmonic
  Locked Loops}. In \bibinfo{booktitle}{\emph{2017 IEEE International
  Conference on Acoustics, Speech and Signal Processing (ICASSP)}}.
  \bibinfo{pages}{191--195}.
\newblock
\urldef\tempurl%
\url{https://doi.org/10.1109/ICASSP.2017.7952144}
\showDOI{\tempurl}


\bibitem[Borkar et~al\mbox{.}(2021)]%
        {borkar2021music}
\bibfield{author}{\bibinfo{person}{Neetish Borkar}, \bibinfo{person}{Shubhra
  Patre}, \bibinfo{person}{Raunak~Singh Khalsa}, \bibinfo{person}{Rohanshhi
  Kawale}, {and} \bibinfo{person}{Priti Chakurkar}.}
  \bibinfo{year}{2021}\natexlab{}.
\newblock \showarticletitle{Music Plagiarism Detection using Audio
  Fingerprinting and Segment Matching}. In \bibinfo{booktitle}{\emph{2021 Smart
  Technologies, Communication and Robotics (STCR)}}. IEEE,
  \bibinfo{pages}{1--4}.
\newblock


\bibitem[Chandna et~al\mbox{.}(2020)]%
        {9053024}
\bibfield{author}{\bibinfo{person}{Pritish Chandna}, \bibinfo{person}{Merlijn
  Blaauw}, \bibinfo{person}{Jordi Bonada}, {and} \bibinfo{person}{Emilia
  Gómez}.} \bibinfo{year}{2020}\natexlab{}.
\newblock \showarticletitle{Content Based Singing Voice Extraction from a
  Musical Mixture}. In \bibinfo{booktitle}{\emph{ICASSP 2020 - 2020 IEEE
  International Conference on Acoustics, Speech and Signal Processing
  (ICASSP)}}. \bibinfo{pages}{781--785}.
\newblock
\urldef\tempurl%
\url{https://doi.org/10.1109/ICASSP40776.2020.9053024}
\showDOI{\tempurl}


\bibitem[Cronin(2016)]%
        {musicCourt}
\bibfield{author}{\bibinfo{person}{C Cronin}.} \bibinfo{year}{2016}\natexlab{}.
\newblock \showarticletitle{Columbia Law School \& UCLA Law Copyright
  Infringement Project}.
\newblock \bibinfo{journal}{\emph{update}} (\bibinfo{year}{2016}).
\newblock


\bibitem[De~Prisco et~al\mbox{.}(2017a)]%
        {de2017music}
\bibfield{author}{\bibinfo{person}{Roberto De~Prisco}, \bibinfo{person}{Antonio
  Esposito}, \bibinfo{person}{Nicola Lettieri}, \bibinfo{person}{Delfina
  Malandrino}, \bibinfo{person}{Donato Pirozzi}, \bibinfo{person}{Gianluca
  Zaccagnino}, {and} \bibinfo{person}{Rocco Zaccagnino}.}
  \bibinfo{year}{2017}\natexlab{a}.
\newblock \showarticletitle{Music plagiarism at a glance: metrics of similarity
  and visualizations}. In \bibinfo{booktitle}{\emph{2017 21st International
  Conference Information Visualisation (IV)}}. IEEE, \bibinfo{pages}{410--415}.
\newblock


\bibitem[De~Prisco et~al\mbox{.}(2016)]%
        {de2016visualization}
\bibfield{author}{\bibinfo{person}{Roberto De~Prisco}, \bibinfo{person}{Nicola
  Lettieri}, \bibinfo{person}{Delfina Malandrino}, \bibinfo{person}{Donato
  Pirozzi}, \bibinfo{person}{Gianluca Zaccagnino}, {and} \bibinfo{person}{Rocco
  Zaccagnino}.} \bibinfo{year}{2016}\natexlab{}.
\newblock \showarticletitle{Visualization of music plagiarism: Analysis and
  evaluation}. In \bibinfo{booktitle}{\emph{2016 20th International Conference
  Information Visualisation (IV)}}. IEEE, \bibinfo{pages}{177--182}.
\newblock


\bibitem[De~Prisco et~al\mbox{.}(2017b)]%
        {de2017computational}
\bibfield{author}{\bibinfo{person}{Roberto De~Prisco}, \bibinfo{person}{Delfina
  Malandrino}, \bibinfo{person}{Gianluca Zaccagnino}, {and}
  \bibinfo{person}{Rocco Zaccagnino}.} \bibinfo{year}{2017}\natexlab{b}.
\newblock \showarticletitle{A computational intelligence text-based detection
  system of music plagiarism}. In \bibinfo{booktitle}{\emph{2017 4th
  International Conference on Systems and Informatics (ICSAI)}}. IEEE,
  \bibinfo{pages}{519--524}.
\newblock


\bibitem[De~Prisco et~al\mbox{.}(2017c)]%
        {de2017fuzzy}
\bibfield{author}{\bibinfo{person}{Roberto De~Prisco}, \bibinfo{person}{Delfina
  Malandrino}, \bibinfo{person}{Gianluca Zaccagnino}, {and}
  \bibinfo{person}{Rocco Zaccagnino}.} \bibinfo{year}{2017}\natexlab{c}.
\newblock \showarticletitle{Fuzzy vectorial-based similarity detection of music
  plagiarism}. In \bibinfo{booktitle}{\emph{2017 IEEE International Conference
  on Fuzzy Systems (FUZZ-IEEE)}}. IEEE, \bibinfo{pages}{1--6}.
\newblock


\bibitem[Dittmar et~al\mbox{.}(2012)]%
        {dittmar2012audio}
\bibfield{author}{\bibinfo{person}{Christian Dittmar}, \bibinfo{person}{Kay~F
  Hildebrand}, \bibinfo{person}{Daniel G{\"a}rtner}, \bibinfo{person}{Manuel
  Winges}, \bibinfo{person}{Florian M{\"u}ller}, {and} \bibinfo{person}{Patrick
  Aichroth}.} \bibinfo{year}{2012}\natexlab{}.
\newblock \showarticletitle{Audio forensics meets music information
  retrieval—a toolbox for inspection of music plagiarism}. In
  \bibinfo{booktitle}{\emph{2012 Proceedings of the 20th European signal
  processing conference (EUSIPCO)}}. IEEE, \bibinfo{pages}{1249--1253}.
\newblock


\bibitem[Doraisamy and R{\"u}ger(2003)]%
        {doraisamy2003robust}
\bibfield{author}{\bibinfo{person}{Shyamala Doraisamy} {and}
  \bibinfo{person}{Stefan R{\"u}ger}.} \bibinfo{year}{2003}\natexlab{}.
\newblock \showarticletitle{Robust polyphonic music retrieval with n-grams}.
\newblock \bibinfo{journal}{\emph{Journal of Intelligent Information Systems}}
  \bibinfo{volume}{21}, \bibinfo{number}{1} (\bibinfo{year}{2003}),
  \bibinfo{pages}{53--70}.
\newblock


\bibitem[Downie et~al\mbox{.}(2008)]%
        {downie2008audio}
\bibfield{author}{\bibinfo{person}{J~Stephen Downie}, \bibinfo{person}{Mert
  Bay}, \bibinfo{person}{Andreas~F Ehmann}, {and} \bibinfo{person}{M~Cameron
  Jones}.} \bibinfo{year}{2008}\natexlab{}.
\newblock \showarticletitle{Audio Cover Song Identification: MIREX 2006-2007
  Results and Analyses.}. In \bibinfo{booktitle}{\emph{ISMIR}}.
  \bibinfo{pages}{468--474}.
\newblock


\bibitem[Du et~al\mbox{.}(2021)]%
        {9414190}
\bibfield{author}{\bibinfo{person}{Xingjian Du}, \bibinfo{person}{Bilei Zhu},
  \bibinfo{person}{Qiuqiang Kong}, {and} \bibinfo{person}{Zejun Ma}.}
  \bibinfo{year}{2021}\natexlab{}.
\newblock \showarticletitle{Singing Melody Extraction from Polyphonic Music
  based on Spectral Correlation Modeling}. In \bibinfo{booktitle}{\emph{ICASSP
  2021 - 2021 IEEE International Conference on Acoustics, Speech and Signal
  Processing (ICASSP)}}. \bibinfo{pages}{241--245}.
\newblock
\urldef\tempurl%
\url{https://doi.org/10.1109/ICASSP39728.2021.9414190}
\showDOI{\tempurl}


\bibitem[Jurafsky and Martin({[n.\,d.]})]%
        {2010speech}
\bibfield{author}{\bibinfo{person}{Daniel Jurafsky} {and}
  \bibinfo{person}{James~H Martin}.} \bibinfo{year}{[n.\,d.]}\natexlab{}.
\newblock \bibinfo{title}{Speech and Language Processing: An Introduction to
  Natural Language Processing, Computational Linguistics, and Speech
  Recognition}.
\newblock
\newblock


\bibitem[Kuhn(1955)]%
        {https://doi.org/10.1002/nav.3800020109}
\bibfield{author}{\bibinfo{person}{H.~W. Kuhn}.}
  \bibinfo{year}{1955}\natexlab{}.
\newblock \showarticletitle{The Hungarian method for the assignment problem}.
\newblock \bibinfo{journal}{\emph{Naval Research Logistics Quarterly}}
  \bibinfo{volume}{2}, \bibinfo{number}{1-2} (\bibinfo{year}{1955}),
  \bibinfo{pages}{83--97}.
\newblock
\urldef\tempurl%
\url{https://doi.org/10.1002/nav.3800020109}
\showDOI{\tempurl}
\showeprint{https://onlinelibrary.wiley.com/doi/pdf/10.1002/nav.3800020109}


\bibitem[Manning and Schütze(1999)]%
        {1999Foundations}
\bibfield{author}{\bibinfo{person}{C.~D. Manning} {and} \bibinfo{person}{H
  Schütze}.} \bibinfo{year}{1999}\natexlab{}.
\newblock \bibinfo{booktitle}{\emph{Foundations of Statistical Natural Language
  Processing}}.
\newblock \bibinfo{publisher}{Foundations of Statistical Natural Language
  Processing}.
\newblock


\bibitem[M{\"u}llensiefen et~al\mbox{.}(2004)]%
        {0cognitive}
\bibfield{author}{\bibinfo{person}{Daniel M{\"u}llensiefen},
  \bibinfo{person}{Klaus Frieler}, {et~al\mbox{.}}}
  \bibinfo{year}{2004}\natexlab{}.
\newblock \showarticletitle{Cognitive adequacy in the measurement of melodic
  similarity: Algorithmic vs. human judgments}.
\newblock \bibinfo{journal}{\emph{Computing in Musicology}}
  \bibinfo{volume}{13}, \bibinfo{number}{2003} (\bibinfo{year}{2004}),
  \bibinfo{pages}{147--176}.
\newblock


\bibitem[Müllensiefen and Pendzich(2009)]%
        {doi:10.1177/102986490901300111}
\bibfield{author}{\bibinfo{person}{Daniel Müllensiefen} {and}
  \bibinfo{person}{Marc Pendzich}.} \bibinfo{year}{2009}\natexlab{}.
\newblock \showarticletitle{Court decisions on music plagiarism and the
  predictive value of similarity algorithms}.
\newblock \bibinfo{journal}{\emph{Musicae Scientiae}} \bibinfo{volume}{13},
  \bibinfo{number}{1\_suppl} (\bibinfo{year}{2009}), \bibinfo{pages}{257--295}.
\newblock
\urldef\tempurl%
\url{https://doi.org/10.1177/102986490901300111}
\showDOI{\tempurl}
\showeprint{https://doi.org/10.1177/102986490901300111}


\bibitem[Pachet et~al\mbox{.}(2006)]%
        {pachet2006cuidado}
\bibfield{author}{\bibinfo{person}{Fran{\c{c}}ois Pachet},
  \bibinfo{person}{Jean-Julien Aucouturier}, \bibinfo{person}{Amaury
  La~Burthe}, \bibinfo{person}{Aymeric Zils}, {and} \bibinfo{person}{Anthony
  Beurive}.} \bibinfo{year}{2006}\natexlab{}.
\newblock \showarticletitle{The cuidado music browser: an end-to-end electronic
  music distribution system}.
\newblock \bibinfo{journal}{\emph{Multimedia Tools and Applications}}
  \bibinfo{volume}{30}, \bibinfo{number}{3} (\bibinfo{year}{2006}),
  \bibinfo{pages}{331--349}.
\newblock


\bibitem[Robine et~al\mbox{.}(2007a)]%
        {10.1145/1290082.1290103}
\bibfield{author}{\bibinfo{person}{Matthias Robine}, \bibinfo{person}{Pierre
  Hanna}, {and} \bibinfo{person}{Pascal Ferraro}.}
  \bibinfo{year}{2007}\natexlab{a}.
\newblock \showarticletitle{Music Similarity: Improvements of Edit-Based
  Algorithms by Considering Music Theory} \emph{(\bibinfo{series}{MIR '07})}.
  \bibinfo{publisher}{Association for Computing Machinery},
  \bibinfo{address}{New York, NY, USA}, \bibinfo{pages}{135–142}.
\newblock
\showISBNx{9781595937780}
\urldef\tempurl%
\url{https://doi.org/10.1145/1290082.1290103}
\showDOI{\tempurl}


\bibitem[Robine et~al\mbox{.}(2007b)]%
        {robine2007adaptation}
\bibfield{author}{\bibinfo{person}{Matthias Robine}, \bibinfo{person}{Pierre
  Hanna}, \bibinfo{person}{Pascal Ferraro}, {and} \bibinfo{person}{Julien
  Allali}.} \bibinfo{year}{2007}\natexlab{b}.
\newblock \showarticletitle{Adaptation of string matching algorithms for
  identification of near-duplicate music documents}. In
  \bibinfo{booktitle}{\emph{Workshop on Plagiarism Analysis, Authorship
  Identification, and Near-Duplicate Detection (PAN07)}}.
  \bibinfo{pages}{37--43}.
\newblock


\bibitem[Sie et~al\mbox{.}(2017)]%
        {siedetecting}
\bibfield{author}{\bibinfo{person}{Mu-Syuan Sie}, \bibinfo{person}{Cheng-Chin
  Chiang}, \bibinfo{person}{Hsiu-Chun Yang}, {and} \bibinfo{person}{Yi-Le
  Liu}.} \bibinfo{year}{2017}\natexlab{}.
\newblock \showarticletitle{DETECTING AND LOCATING PLAGIARISM OF MUSIC MELODIES
  BY PATH EXPLORATION OVER ABinary MASK}. In \bibinfo{booktitle}{\emph{CS \& IT
  Conference Proceedings}}, Vol.~\bibinfo{volume}{7}. CS \& IT Conference
  Proceedings.
\newblock


\bibitem[Tversky(1988)]%
        {1988Features}
\bibfield{author}{\bibinfo{person}{A. Tversky}.}
  \bibinfo{year}{1988}\natexlab{}.
\newblock \showarticletitle{Features of similarity.}
\newblock \bibinfo{journal}{\emph{Readings in Cognitive Science}}
  \bibinfo{volume}{84}, \bibinfo{number}{4} (\bibinfo{year}{1988}),
  \bibinfo{pages}{290--302}.
\newblock


\bibitem[Uhlich et~al\mbox{.}(2015)]%
        {7178348}
\bibfield{author}{\bibinfo{person}{Stefan Uhlich}, \bibinfo{person}{Franck
  Giron}, {and} \bibinfo{person}{Yuki Mitsufuji}.}
  \bibinfo{year}{2015}\natexlab{}.
\newblock \showarticletitle{Deep neural network based instrument extraction
  from music}. In \bibinfo{booktitle}{\emph{2015 IEEE International Conference
  on Acoustics, Speech and Signal Processing (ICASSP)}}.
  \bibinfo{pages}{2135--2139}.
\newblock
\urldef\tempurl%
\url{https://doi.org/10.1109/ICASSP.2015.7178348}
\showDOI{\tempurl}


\bibitem[Uitdenbogerd and Zobel(1999)]%
        {inproceedings}
\bibfield{author}{\bibinfo{person}{Alexandra Uitdenbogerd} {and}
  \bibinfo{person}{Justin Zobel}.} \bibinfo{year}{1999}\natexlab{}.
\newblock \showarticletitle{Melodic matching techniques for large music
  databases}.
\newblock \bibinfo{journal}{\emph{Proceedings of the ACM International
  Multimedia Conference \& Exhibition}}, \bibinfo{pages}{57--66}.
\newblock
\urldef\tempurl%
\url{https://doi.org/10.1145/319463.319470}
\showDOI{\tempurl}


\bibitem[Velardo et~al\mbox{.}(2016)]%
        {velardo2016symbolic}
\bibfield{author}{\bibinfo{person}{Valerio Velardo}, \bibinfo{person}{Mauro
  Vallati}, {and} \bibinfo{person}{Steven Jan}.}
  \bibinfo{year}{2016}\natexlab{}.
\newblock \showarticletitle{Symbolic melodic similarity: State of the art and
  future challenges}.
\newblock \bibinfo{journal}{\emph{Computer Music Journal}}
  \bibinfo{volume}{40}, \bibinfo{number}{2} (\bibinfo{year}{2016}),
  \bibinfo{pages}{70--83}.
\newblock


\bibitem[Wang et~al\mbox{.}(2020)]%
        {wang2020pop909}
\bibfield{author}{\bibinfo{person}{Ziyu Wang}, \bibinfo{person}{Ke Chen},
  \bibinfo{person}{Junyan Jiang}, \bibinfo{person}{Yiyi Zhang},
  \bibinfo{person}{Maoran Xu}, \bibinfo{person}{Shuqi Dai},
  \bibinfo{person}{Xianbin Gu}, {and} \bibinfo{person}{Gus Xia}.}
  \bibinfo{year}{2020}\natexlab{}.
\newblock \showarticletitle{Pop909: A pop-song dataset for music arrangement
  generation}.
\newblock \bibinfo{journal}{\emph{arXiv preprint arXiv:2008.07142}}
  (\bibinfo{year}{2020}).
\newblock


\bibitem[Wu and Yang(2021)]%
        {wu2021musemorphose}
\bibfield{author}{\bibinfo{person}{Shih-Lun Wu} {and} \bibinfo{person}{Yi-Hsuan
  Yang}.} \bibinfo{year}{2021}\natexlab{}.
\newblock \showarticletitle{MuseMorphose: Full-Song and Fine-Grained Music
  Style Transfer with One Transformer VAE}.
\newblock \bibinfo{journal}{\emph{arXiv preprint arXiv:2105.04090}}
  (\bibinfo{year}{2021}).
\newblock


\end{thebibliography}










\end{document}